\documentclass[12 pt]{article} 
\usepackage{times,bm}
\usepackage[affil-it,]{authblk}
\usepackage[pagebackref=false,colorlinks,linkcolor=blue,citecolor=blue]{hyperref}
\usepackage{geometry}
\usepackage{graphicx}
\usepackage{cite}
\usepackage{amsmath}
\usepackage{amsfonts}
\usepackage{amssymb}
\usepackage{color}
\usepackage{float}
\usepackage{multirow,multicol}
\usepackage{tabulary}
\usepackage[flushleft]{threeparttable}
\usepackage{pgfplots,subfigure}
\usepackage{xcolor}
\numberwithin{equation}{section}
\usepackage{tikz}
\usepgflibrary{arrows}
\usetikzlibrary{shapes.callouts}
\tikzset{  
	level/.style   = { thick, },
	connect/.style = { dotted, red   },
	notice/.style  = { draw, rectangle callout, callout relative pointer={#1} },
	label/.style   = { text width=2cm }
}

\usepackage{letltxmacro}
\makeatletter
\let\oldr@@t\r@@t
\def\r@@t#1#2{%
	\setbox0=\hbox{$\oldr@@t#1{#2\,}$}\dimen0=\ht0
	\advance\dimen0-0.2\ht0
	\setbox2=\hbox{\vrule height\ht0 depth -\dimen0}%
	{\box0\lower0.4pt\box2}}
\LetLtxMacro{\oldsqrt}{\sqrt}
\renewcommand*{\sqrt}[2][\ ]{\oldsqrt[#1]{#2}}
\makeatother

\begin{document}

	\title{ Reply to "\textit{Comment on "Effects of cosmic-string framework on the
				thermodynamical properties of anharmonic oscillator using
				the ordinary statistics and the	q-deformed superstatistics	approaches"}"}

	\author[1]{\textit{Hadi Sobhani}%
		\thanks{Electronic address: hadisobhani8637@gmail.com (Corresponding author)}}
	
	\author[1]{\textit{Hassan Hassanabadi}}
	\affil[1]{{\small Faculty of Physics, Shahrood University of Technology, Shahrood, Iran \\ P. O. Box: 3619995161-316.}}
	
		\author[2]{\textit{Won Sang Chung}}
	\affil[2]{{\small Department of Physics and Research Institute of Natural Science,
			College of Natural Science, Gyeongsang National University, Jinju 52828, Korea }}
		
	\date{}
	\maketitle
	\thispagestyle{empty}

	\begin{abstract}
		\textit{
		In this paper, we show the detail of our recent paper \emph{Effects of  cosmic-string framework on the
			thermodynamical properties of an anharmonic oscillator using the ordinary statistics and the	q-deformed superstatistics	approaches}
		published in the Eur. Phys. J. C. Actually, we prepare this comment against the comment prepared by \textbf{Francisco A. Cruz Neto} and \textbf{Luis B. Castro}.
		}
	\end{abstract}
	
	\begin{small}
		
	\end{small}
	
	\newpage

\section{Introduction}

Recently, we have published the paper "\emph{Effects of cosmic-string framework on the	thermodynamical properties of anharmonic oscillator using
	the ordinary statistics and the	q-deformed superstatistics	approaches}" \cite{ours} in which we discussed the thermodynamical properties of the anharmonic canonical ensemble within
the cosmic-string framework. We were aware that the authors \textbf{Francisco A. Cruz Neto} and \textbf{Luis B. Castro} commented on out paper and claimed that we took a wrong way in our paper \cite{comment}. In the following, we show the detail of our derivation and discuss the results.

\section{Detail of the Derivations}

Before we start showing detail of the derivations, we'd like to thank the authors \textbf{Francisco A. Cruz Neto} and \textbf{Luis B. Castro}, because when we checked our paper, we found out that there were some typo-errors that we want to correct them in this paper but we emphasize that they can change the physical results. We start from 
\begin{align}
\label{2-1}
\frac{d^2 f (y)}{d y^2} +  \left( \frac{1}{4 y^2} + \frac{\frac{1}{4} (\varepsilon - k^2)}{y} - \frac{1}{4}(a_v + l^2 \alpha^2)   - \frac{b_v}{4} y -  \frac{c_v}{4} y^2 \right) f (y)=0.
\end{align}
where the power $2$ of the first term in the parentheses was missed in our paper. Since we knew that such differential equation can be solved with the help of bi-confluent Heun functions, we follow the common method used in many  quantum problems and consider the solution in the form of 
\begin{align}
\label{2-2}
f(y) = y^{A_1} \exp [y( B_1 + D_1 y)]F(y),
\end{align}
where the constant $A_1, B_1$ and $D_1$ shall be determined. Determination of these constant will be done knowing original bi-confluent Heun differential equation
\begin{equation}
\label{2-3}
y''(x)+ \left( \frac{1+\alpha'}{x} - \beta' -2x \right) y'(x) + \left( \gamma' - \alpha' - 2 - \frac{\delta'+(1+\alpha')\beta'}{2x}\right) y(x)=0.  
\end{equation}  
in which a minus sign was missed in our paper and the prime means derivative with respect the independent variable. Taking differentiating for two times from Eq. \eqref{2-2} and substituting the results into Eq. \eqref{2-1} we have
\begin{align}
\label{2-4}
\begin{gathered}
F''(y) + \left( {\frac{{2A}}{y} + 2B + 4Dy} \right)F'(y) +  \hfill \\
F(y)\left( \begin{gathered}
\frac{{{A^2} - A + \tfrac{1}{4}}}{{{y^2}}} + \frac{{2AB + \tfrac{1}{4}\left( {\varepsilon  - {k^2}} \right)}}{y} - \frac{1}{4}\left( {{a_v} + {l^2}{\alpha ^2}} \right) \hfill \\
+ {B^2} + 2D(1 + 2A) + \left( { - \frac{{{b_v}}}{4} + 4BD} \right)y + \left( { - \frac{c}{4} + 4{D^2}} \right){y^2} \hfill \\ 
\end{gathered}  \right) = 0. \hfill \\ 
\end{gathered} 
\end{align} 
Comparing Eq. \eqref{2-4} and \eqref{2-3}, we find out that  determination of the constants $A_1, B_1$ and $D_1$ will be done by removing the coefficients of the $y^2, y$ and $y^{-2}$. Knowing this point, we obtain three equations for those constants
\begin{align}
\label{2-5}
A^2-A+\frac{1}{4}=0 &\Rightarrow A=\frac{1}{2}, \\
\label{2-6}
- \frac{{{b_v}}}{4} + 4BD=0 &\Rightarrow B = - \frac{b_v}{8},\\
\label{2-7}
- \frac{c}{4} + 4 D^2 =0 &\Rightarrow D=-\frac{\sqrt{c_v}}{4},
\end{align}
where we only choose the signs that have physical meaning. If we note Eqs. \eqref{2-4} and \eqref{2-3}, we see that if we want to reach to the term $-2x$ in Eq. \eqref{2-3}, we have two different choices:
\begin{itemize}
	\item 
	This first choice is using a variable changing as the authors of Ref. \cite{comment} used in their paper. 
	
	\item 
	The second choice is using an approximation (we made use of setting a numerical value for the potential constant) which brings us a better physical interpretation. Such a selection is very common in solving quantum mechanical problems because it satisfies the postulates of quadratically integrability of the wave function which guarantees that the integral of the square of the absolute value of the wave function be finite \cite{q1,q2,q3,q4,q5}. We take this choice thus we have 
	\begin{align}
	\label{2-8}
	-\frac{1}{2} = - \frac{\sqrt{c_\nu}}{4} \rightarrow c_\nu=4.
	\end{align}
\end{itemize}
So, using Eqs. \eqref{2-5}, \eqref{2-6}, \eqref{2-7} and \eqref{2-8}, we rewrite Eq. \eqref{2-4} as
\begin{align}
\label{2-9}
F''(y) + \left( {\frac{1}{y} - \frac{{{b_v}}}{4} - 2y} \right)F'(y) + \left( \begin{gathered}
\frac{{ - \frac{{{b_v}}}{8} + \frac{1}{4}(\varepsilon  - {k^2})}}{y} - \frac{1}{4}\left( {{a_v} + {l^2}{\alpha ^2}} \right) - 2 + \frac{b_v ^2}{64} \hfill \\ 
\end{gathered}  \right)F(y) = 0.
\end{align}
Comparing between Eqs. \eqref{2-9} and \eqref{2-3} we have 
\begin{align}
\label{2-10}
&\alpha'=0, \\
\label{2-11}
&\beta'=\frac{b_v}{4}, \\
\label{2-12}
&\gamma'= \frac{b_v ^2}{64} - \frac{1}{4} (a_v ^2 + l^2 \alpha^2 ), \\
\label{2-13}
&\delta'=\frac{1}{2}(k^2 - \varepsilon),
\end{align}
where we had a typo error in $\gamma$ parameter of Heun function in our paper. Finally, the solution of Eq. \eqref{2-9} is written in terms of Heun functions
\begin{align}
\label{2-14}
F(y)=H_b(\alpha', \beta', \gamma', \delta',y).
\end{align}

After determination of the wave function, it is the time for deriving energy eigenvalue relation. To obtain the energy eigenvalue relation of the considered system, we should use the series form of the wave function as what is done for a simple harmonic oscillator or some other famous problems discussed in many quantum mechanics books \cite{q1,q2,q3,q4,q5}.

For the sake of simplicity, let us check this points for Eq. \eqref{2-3}. Considering the series form of the solution we have 
\begin{align}
\label{2-15}
y(x) &= \sum_{n=0}^{\infty} a_n x^n,  \\
\label{2-16}
y'(x) &= \sum_{n=1}^{\infty} n a_n x^{n-1},  \\
\label{2-17}
y''(x) &= \sum_{n=2}^{\infty} n(n-1) a_n x^{n-2}. 
\end{align}
Substituting Eqs. \eqref{2-15}, \eqref{2-16} and \eqref{2-17} into \eqref{2-3} yields
\begin{align}
\label{2-18}
\begin{gathered}
\sum\limits_{n = 0}^\infty  {(n + 1)} (n + 2){a_{n + 2}}{x^n} + (1 + \alpha' )\sum\limits_{n = 0}^\infty  {(n + 1)} {a_{n + 1}}{x^{n - 1}} - \beta' \sum\limits_{n = 0}^\infty  {(n + 1)} {a_{n + 1}}{x^n} \hfill \\
- 2\sum\limits_{n = 0}^\infty (n+1) {{a_{n + 1}}} {x^{n + 1}} + (\gamma'  - \alpha'  - 2)\sum\limits_{n = 0}^\infty  {{a_n}{x^n}  } -\frac{{\delta'  + (1 + \alpha' )\beta' }}{2}\sum\limits_{n = 0}^\infty  {{a_n}} {x^{n - 1}} = 0. \hfill \\ 
\end{gathered} 
\end{align} 
From Eq.\eqref{2-18}, we want to find coefficient of $x^{n+1}$ which results
\begin{align}
\label{2-19}
\begin{gathered}
{a_{n + 3}}(n + 3)(n + 3 + \alpha ') + {a_{n + 2}}\left( { - \beta '(n + 2) - \frac{{\delta ' + (1 + \alpha ')\beta '}}{2}} \right) \hfill \\
+ {a_{n + 1}}(\gamma ' - \alpha ' - 4 - 2n) = 0. \hfill \\ 
\end{gathered} 
\end{align}
The series should be cut off after large enough value of $n$. It means that we should  equal the coefficients of $a_{n+3}$ to zero. It requires that the coefficients of $a_{n+2}$ and $a_{n+1}$ should be equal to zero. Therefore we find two constraints
\begin{align}
\label{2-20}
&{ - \beta '(n + 2) - \frac{{\delta ' + (1 + \alpha ')\beta '}}{2}}=0, \\
\label{2-21}
& \gamma ' - \alpha ' - 4 - 2n=0.
\end{align}
which are not in contrast with Refs. \cite{4,5,6,7,8,9}. Using these constraints, we can obtain the energy eigenvalue and a constraint on a potential constant
\begin{align}
\label{2-22}
\varepsilon &= \frac{b_v}{2} (2n+5)  + k^2, \\
\label{2-23}
b_v &= \pm 4 \sqrt{a_v ^2 + \alpha^2 l^2 + 8n + 16}.
\end{align} 
It is seen that in the new expression of the constraint on the potential constant, we see a numerical difference which obviously does not change the physical interpretation. If one uses the positive sign of Eq. \eqref{2-23}, it is seen that no physical changes happen in the results. If we did not use the minus sign, it did not mean that is not physical. 

The authors of Ref. \cite{comment} only discussed from the mathematical point of view and discussed the more different solution to the results but it does imply that we deny the existence of the general cases. We used some physical considerations which are so common in solving of quantum mechanical problems and made the problem easier such that more reader can understand it without losing the generality and missing a physical interpretation. It was shown that it does not affect the physical interpretation of our results.




\begin{thebibliography}{99}
	
\bibitem{ours}
H. Sobhani, H. Hassanabadi, W. S. Chung, \textit{Eur.Phys. J. C} \textbf{78} (2)  106 (2018). 

\bibitem{comment}
Francisco A. Cruz Neto, Luis B. Castro, 	arXiv:1804.03012 (2018).

\bibitem{q1} 
S. Gasiorowicz, \textit{Quantum Physics, 3rd Edition}, ISBN: 978-0-471-05700-0, (2003)

\bibitem{q2}

David J. Griffiths, \textit{Introduction to Quantum Mechanics}, (Pearson Prentice Hall) (2nd edition), (2010)

\bibitem{q3}
E. Merzbacher, \textit{Quantum Mechanics}  (John Wiley and Sons, Inc. New York) (1955)

\bibitem{q4} 
Walter Greiner, \textit{Quantum Mechanics, An Introduction} (Springer-Verlag Berlin Heidelberg) (2001)

\bibitem{q5}
Nouredine Zettili, \textit{Quantum Mechanics: Concepts and Applications}, (Wiley) (2nd edition) (2009)

\bibitem{4}
L.B. Castro, \textit{Phys. Rev. C} \textbf{86}  052201 (2012). 

\bibitem{5}
E.R. Figueiredo Medeiros, E.R. Bezerra de Mello, \textit{Eur. Phys. J. C} \textbf{72} (6)  2051 (2012). 

\bibitem{6}
K. Bakke, F. Moraes, \textit{Phys. Lett. A} \textbf{376} (45)  2838 (2012). 

\bibitem{7}
K. Bakke, \textit{Ann. Phys. (N.Y.)} \textbf{341}  86 (2014). 

\bibitem{8}
K. Bakke, C. Furtado, \textit{Ann. Phys. (N.Y.)} \textbf{355}  48 (2015). 

\bibitem{9}
R. L. L. Vit\'{o}ria, C. Furtado, K. Bakke, \textit{Eur. Phys. J. C} \textbf{78} (1)  44 (2018).


\end{thebibliography}
\end{document}